\documentclass[twocolumn,aps,prd,preprintnumbers,nofootinbib,superscriptaddress]{revtex4-2}
\usepackage{graphicx}
\usepackage{amsmath}
\usepackage{amssymb}
\usepackage{amsfonts}
\usepackage{hyperref}
\usepackage{url}
\usepackage{color}
\usepackage{bm}
\usepackage[dvipsnames]{xcolor}

\newcommand{\be}{\begin{equation}}
\newcommand{\ee}{\end{equation}}
\newcommand{\ba}{\begin{eqnarray}}
\newcommand{\ea}{\end{eqnarray}}
\newcommand{\nn}{\nonumber}

\renewcommand{\[}{\begin{equation}}
\renewcommand{\]}{\end{equation}}

\def\be{\begin{equation}}
\def\ee{\end{equation}}
\def\bea{\begin{eqnarray}}
\def\eea{\end{eqnarray}}
\def\eqi{\begin{equation}}
\def\eqf{\end{equation}}
\def\eqia{\begin{eqnarray}}
\def\eqfa{\end{eqnarray}}
\def\lcdm{$\Lambda$CDM }

\def\lcdm{$\Lambda$CDM }

\usepackage{array}
\makeatother

\begin{document}

\preprint{IFT-UAM/CSIC-24-173}

\title{Comparative analysis of the NANOgrav Hellings-Downs as a window into new physics}

\author{Rub\'{e}n Arjona}
\email{rarjona@ucm.es}
\affiliation{Instituto de F\'isica Te\'orica UAM-CSIC, Universidad Auton\'oma de Madrid,
Cantoblanco, 28049 Madrid, Spain}

\author{Savvas Nesseris}
\email{savvas.nesseris@csic.es}
\affiliation{Instituto de F\'isica Te\'orica UAM-CSIC, Universidad Auton\'oma de Madrid,
Cantoblanco, 28049 Madrid, Spain}

\author{Sachiko Kuroyanagi}
\email{sachiko.kuroyanagi@csic.es}
\affiliation{Instituto de F\'isica Te\'orica UAM-CSIC, Universidad Auton\'oma de Madrid,
Cantoblanco, 28049 Madrid, Spain}
\affiliation{Department of Physics and Astrophysics, Nagoya University, Nagoya, 464-8602, Japan}

\date{\today}

\begin{abstract}
Pulsar timing array (PTA) experiments have recently provided strong evidence for the signal of the stochastic gravitational wave background (SGWB) in the nHz-frequency band. These experiments have shown a statistical preference for the Hellings-Downs (HD) correlation between pulsars, which is widely regarded as a definitive signature of the SGWB. Using the NANOGrav 15-year dataset, we perform a comparative Bayesian analysis of four different models that go beyond the standard cosmological framework and influence the overlap reduction function. Specifically, we analyze ultralight vector dark matter (DM), spin-2 ultralight DM, massive gravity, and a folded non-Gaussian component to the SGWB. We find that the spin-2 ultralight DM and the massive gravity model are statistically equivalent to the HD prediction, and there is weak evidence in favor of the non-Gaussian component and the ultralight vector DM model. We also perform a non-parametric test using the Genetic Algorithms, which suggests a weak deviation from the HD curve. However, improved data quality is required before drawing definitive conclusions.
\end{abstract}

\maketitle

\section{Introduction\label{sec:intro}}
Along with black holes and the expansion of the Universe, the presence of gravitational radiation is one of the central predictions of Einstein’s general theory of relativity (GR)~\cite{einstein1918gravitationswellen}. The discovery of the binary pulsar PSR B1913+16~\cite{hulse1975discovery} and the later observation of its orbital decay, along with similar findings from other binary pulsars, have provided compelling evidence for the existence of gravitational waves (GWs)~\cite{weisberg2016relativistic}. These discoveries have sparked a global effort to detect GWs directly, primarily through the use of kilometer-scale laser interferometers like the LIGO, Virgo, and KAGRA detectors~\cite{aasi2015advanced,acernese2014advanced,KAGRA:2018plz}.

Another exciting avenue for exploring the existence of GWs is pulsar timing arrays (PTAs). PTA observations provide a powerful method for investigating the stochastic gravitational wave background (SGWB)~\cite{Sazhin:1978myk,Detweiler:1979wn,hellings1983upper,Jenet:2005pv}, which is a type of GW signal characterized by random fluctuations originating from all directions. As a GW propagates through spacetime, it induces perturbations in the arrival times of pulses from pulsars. These perturbations exhibit distinct spatial correlation patterns, known as Hellings-Downs (HD) correlations, which arise from the quadrupolar nature of GWs~\cite{hellings1983upper,romano2023answers}.

Recently, PTA collaborations such as the North American Nanohertz Observatory for Gravitational Waves (NANOGrav)~\cite{arzoumanian2009nanograv,agazie2306nanograv}, the European Pulsar Timing Array (EPTA)~\cite{balaji2023scalar,antoniadis2023second}, the Parkes Pulsar Timing Array (PPTA)~\cite{goncharov2107evidence,reardon2306search}, and the Chinese Pulsar Timing Array~\cite{xu2306searching} have
reported strong evidence for a correlated low-frequency stochastic process in pulsar timing residuals. By analyzing correlations in the precise timing of signals from tens of millisecond pulsars, they have found indications of the HD correlations. Furthermore, the amplitude and spectral index of the SGWB power spectrum have been measured. These findings suggest that the observed SGWB may originate from the binaries of supermassive black holes or from cosmological GWs~\cite{NANOGrav:2023hvm,epta2023second,burke2019astrophysics,winkler2024origin}. As the signal-to-noise ratio of these observations improves over time, they are becoming an increasingly compelling tool for exploration.

The PTA observation involve two key components: the power spectrum and the overlap reduction function. The power spectrum, which characterizes how the energy density of the SGWB varies with frequency, is often used to explore the origin of the SGWB. In parallel, the overlap reduction function is an independent observational measurement that can be used to probe new physics. It characterizes the spatial dependence of the correlation as a function of the angular separation between pulsar pairs. In the context of GR, for an isotropic, unpolarized, and Gaussian background, the overlap reduction function is known as the HD curve, which is expressed as~\cite{hellings1983upper,romano2023answers}
\begin{equation}\label{eq:HD}
\begin{aligned}
\Gamma_{\mathrm{HD}}(\xi)= & ~\frac{1}{2}-\frac{1}{4}\left(\frac{1-\cos \xi}{2}\right)+ \\
& +\frac{3}{2}\left(\frac{1-\cos \xi}{2}\right) \ln \left(\frac{1-\cos \xi}{2}\right),
\end{aligned}
\end{equation}
where $\xi$ is the separation angle between the pulsar pair. It has been shown that this can be derived also by very general symmetry arguments~\cite{kehagias2024pta}.

The HD curve can be modified by relaxing the assumptions or by considering new sources that induce correlation patterns in the timing residuals. Driven by the exciting discoveries from PTA experiments, there has been active discussion about potential modifications to the HD curve for various mechanisms. Although the data is currently too noisy to fully reconstruct the HD curve, it is still valuable to search for potential deviations from the GR prediction. In this paper, we perform a comparative Bayesian analysis of four different models predicting distinct overlap reduction functions, using the NANOGrav 15-year data set. These models include: ultralight vector dark matter (DM)~\cite{omiya2023hellings}, spin-2 ultralight DM~\cite{cai2024angular}, massive gravity~\cite{liang2021detecting}, and a folded non-Gaussian component to the SGWB~\cite{jiang2024search}. Additionally, we present a non-parametric approach to reconstruct the overlap reduction function from the data using Genetic Algorithms (GA). Both approaches suggest a potential deviation from the HD curve.

The outline of our paper is as follows: In Sec.\ref{sec:models}, we describe the four models that modify the HD curve. In Sec.\ref{sec:analysis}, we perform Bayesian comparison using the data from the NANOGrav 15-year data set. We also provide GA fitting of the overlap reduction function. Finally, we conclude in Sec.~\ref{sec:conclusions}.

\section{The models \label{sec:models}}

\subsection{Ultralight vector dark matter}
Ultralight DM is a model in which DM is represented by the coherent oscillation of an ultralight bosonic field, with its frequency determined by the mass of the field. These oscillations result in fluctuations of the gravitational potential, which in turn affect the timing residuals of pulses emitted by pulsars~\cite{khmelnitsky2014pulsar}. 

In Ref.~\cite{omiya2023hellings}, the authors show that ultralight vector DM can cause a deformation of the HD  curve correlation at the frequency $f=\mu/\pi$ when the DM mass falls within the range $10^{-24} \mathrm{eV} \lesssim \mu \lesssim 10^{-23} \mathrm{eV}$. For this mass range, the oscillation frequency of the vector field falls within the nanohertz band, allowing PTA observations to probe a distinct signature of ultralight DM with spin-1 matter. It has been shown that vector DM generates a combination of monopole and quadrupole angular correlation patterns and leads to the deformation in the HD correlation. The effect is more pronounced for smaller DM masses, but PTA observations are sensitive only down to  $10^{-24} \mathrm{eV}$ due to the limited frequency range they can probe. For higher DM masses, the quadrupole contribution dominates, and the original HD curve is preserved.

The overlap reduction function $\Gamma_{\mathrm{eff}}$ within the frequency band which includes $2\pi f=2\mu$ can be written as
\begin{equation}
\small{
\Gamma_{\mathrm{eff}}(\xi,\mu)=\frac{\Phi_{\mathrm{GW}}(\mu / \pi)}{\Phi_{\mathrm{GW}}(\mu / \pi)+\Phi_{\mathrm{DM}}}\left[\Gamma_{\mathrm{HD}}(\xi)+\frac{\Phi_{\mathrm{DM}}}{\Phi_{\mathrm{GW}}(\mu / \pi)} \Gamma_{\mathrm{DM}}(\xi)\right]},
\end{equation}
where the normalization is selected such that $\Gamma_{\mathrm{eff}}(\xi=0)=1/2$ and $\Gamma_{\mathrm{HD}}(\xi)$ is the HD pattern due to a SGWB. The terms $\Phi_{\mathrm{GW}}(\mu / \pi)$ and $\Phi_{\mathrm{DM}}$ are specifically defined in Eqs.~(37) and (38) of Ref.~\cite{omiya2023hellings}. For the sake of self-consistency in this paper, we provide the detailed expressions for $\{\Phi_{\mathrm{GW}}(\mu / \pi),\Phi_{\mathrm{DM}}\}$ in Appendix \ref{app:vector}. In our Bayes comparison analysis, the DM mass $\mu$ is treated as a free parameter. For this model, we can recover the original HD curve when $\mu\rightarrow \infty$.

\subsection{Spin-2 ultralight dark matter}
The spin-2 ultralight DM has attracted significant interest in recent years~\cite{aoki2016massive,babichev2016bigravitational,marzola2018oscillating,manita2023spin}. The spin properties of ultralight DM are crucial in determining its angular correlation on pulsar timing residuals. For instance, scalar ultralight DM affects pulsars uniformly in all directions~\cite{khmelnitsky2014pulsar}, whereas vector ultralight DM exhibits highly anisotropic behavior~\cite{nomura2020pulsar}. The theoretical foundation of spin-2 ultralight DM arises from bimetric theory~\cite{hassan2012bimetric}. 

The overlap reduction function $\Gamma_{\mathrm{eff}}$ of spin-2 ultralight DM at frequency $f_m=m/2\pi$, where $m$ is the mass of the DM particle, can be written as
\begin{equation}
\begin{aligned}
\Gamma_{\mathrm{eff}}(\xi,m,\alpha)= & \frac{\Phi_{\mathrm{GW}}(m / 2 \pi)}{\Phi_{\mathrm{GW}}(m / 2 \pi)+\Phi_{\mathrm{DM}}(\alpha)} \Gamma_{\mathrm{HD}}(\xi)+ \\
& \frac{\Phi_{\mathrm{DM}}(\alpha)}{\Phi_{\mathrm{GW}}(m / 2 \pi)+\Phi_{\mathrm{DM}}(\alpha)} \Gamma_{\mathrm{DM}}(\xi),
\end{aligned}
\end{equation}
where the normalization is selected such that $\Gamma_{\mathrm{eff}}(\xi=0)=1/2$ and $\Gamma_{\mathrm{HD}}(\xi)$ is the HD pattern due to a SGWB. Here, $\alpha$ is the parameter that corresponds to the coupling between the ultralight DM field and matter~\cite{Armaleo:2020yml}. The terms $\Phi_{\mathrm{GW}}(\mu / \pi)$ and $\Phi_{\mathrm{DM}}(\alpha)$ are specifically defined in Eqs.~(42) and (44) of Ref.~\cite{cai2024angular}. Varying the parameters $\alpha$ and $m$ results in different distortions of the HD curve. We provide the detail expressions of $\{\Phi_{\mathrm{GW}}(m/ 2\pi),\Phi_{\mathrm{DM}}(\alpha)\}$ in Appendix \ref{app:spin2}. In our Bayes comparison analysis, the parameters $m$ and $\alpha$ are treated as free parameters. For this model, we can recover the original HD curve when $m\rightarrow 0$ and $\alpha \rightarrow 0$.

\subsection{Massive Gravity}
In a massive gravity model, additional polarization modes (tensor, vector, and scalar) must be considered, along with corrections arising from the mass of the graviton~\cite{liang2021detecting,Wu:2023rib}. In our analysis, we follow Ref.~\cite{liang2021detecting} where the authors consider the specific case of the ghost-free massive gravity. 
The full two-point correlation function incorporates contributions from all polarization modes. If distinguishing between these modes is not feasible, the signal observed by a PTA would be interpreted as originating from the tensor mode. Under this assumption, the effective overlap reduction function can be written as
\begin{equation}\label{eq:MG}
\Gamma_{\mathrm{eff}}(\xi,A, \Omega)=\Gamma_T(\xi,A)+\Omega\cdot \Gamma_V(\xi,A)+\Omega \cdot \Gamma_S(\xi,A),
\end{equation}
where $\Omega$ is the power spectrum which encodes its frequency dependence, while the overlap reduction function $(\Gamma_T,\Gamma_V,\Gamma_S)$ describe the angular dependence of the tensor, vector, and scalar modes, respectively. For simplicity, we will treat $\Omega$ as a frequency-independent quantity.\footnote{It is important to note that different polarization modes exhibit distinct frequency dependencies, meaning that $\Omega$ is, in principle, a function of frequency. However, within the sensitivity range of current PTA observations, the effect of frequency dependence would  be minimal. Therefore, we approximate $\Omega$ as frequency-independent, following the approach of Ref.~\cite{liang2021detecting}.} The parameter $A$ is related to the magnitude of the longitudinal mode, defined as $A=\frac{|\boldsymbol{k}|}{k_0}$, and the dispersion relation is $k_0^2=|\boldsymbol{k}|^2+m^2$ where $m$ is the mass of the graviton. We provide the detail expressions of $(\Gamma_T,\Gamma_V,\Gamma_S)$ in Appendix \ref{app:massive}. In our Bayes comparison analysis, the parameters $A$ and $\Omega$ are treated as free variables. We can recover the regular HD curve by setting $\Omega \rightarrow 0$ and $A\rightarrow 1$. 

Note that Eq.~\eqref{eq:MG} is not the same as Eq.(32) in~\cite{liang2021detecting} which is expressed as
\begin{equation}
\Gamma_{\rm eff}=\Gamma_{T}+\Gamma_{V}\frac{\Omega_V}{\Omega_T}\frac{\beta_T}{\beta_V}+\Gamma_{S}\frac{\Omega_S}{\Omega_T}\frac{\beta_T}{\beta_S}.
\end{equation} 
where $\beta_S=\beta_V=\beta_T=3/(4\pi)$ are normalization factors. In our approach, to limit the number of free parameters to two, we assume the power spectra of the scalar and vector modes to be identical, $\Omega_S = \Omega_V$. This choice maximizes the contributions from both scalar and vector modes, allowing them to be parameterized by a single parameter, $\Omega$.\footnote{This strategy is inspired by the approach taken in Ref.~\cite{liang2021detecting}, as discussed in their comments below Figure 8.}

\subsection{Non-Gaussian component to SGWB}
Different approaches have been proposed to detect non-Gaussianity of the SGWB using PTA experiments~\cite{tsuneto2019searching,powell2020probing,tasinato2022gravitational}. Some methods target higher-order correlations between pulsars, although these are challenging to detect and computationally expensive. In Ref.~\cite{tasinato2022gravitational}, it has been suggested that it might be possible to search for non-Gaussianity by examining its non-linear effect on the overlap reduction function. 

In this paper, we adopt the methodology outlined in Ref.~\cite{jiang2024search}, which focuses on a folded non-Gaussian component in the SGWB. In the unpolarized case, the amplitude of the trispectrum is characterized by a single parameter $\alpha$. Their work provides the following overlap reduction function
\small{
\begin{equation}
\begin{aligned}
&\Gamma\left(\zeta_{a b}\right) \propto  \\
& \sum_{A=+, \times}\int_{S^2} \mathrm{~d} \hat{\mathbf{n}}\,\Bigg\{ \frac{\bar{E}_a^A \bar{E}_b^A}{\left(1+\hat{\mathbf{n}}_a \cdot \hat{\mathbf{n}}\right)\left(1+\hat{\mathbf{n}}_b \cdot \hat{\mathbf{n}}\right)} \\
&+ \frac{4 \alpha\left[\frac{9}{16} \bar{E}_a^A \bar{E}_a^A \bar{E}_b^A \bar{E}_b^A+\frac{5}{8}\left(\bar{E}_a^A \bar{E}_a^A \bar{E}_a^A \bar{E}_b^A+\bar{E}_a^A \bar{E}_b^A \bar{E}_b^A \bar{E}_b^A\right)\right]}{\left(1+\hat{\mathbf{n}}_a \cdot \hat{\mathbf{n}}\right)\left(1+\hat{\mathbf{n}}_b \cdot \hat{\mathbf{n}}\right)}\Bigg\},
\end{aligned}
\end{equation}
}
\noindent where $\bar{E}_a^A=\mathbf{e}_{i j}^A \,\hat{\mathbf{n}}_a^i \,\hat{\mathbf{n}}_a^j$, and $\mathbf{e}_{i j}^A$ is the basis for the two kinds of polarization tensors. We provide the detailed expressions of $\bar{E}_a^A$ in Appendix~\ref{app:nonG}. In our Bayes comparison analysis, the parameter $\alpha$ is treated as a free parameter, and we recover the original HD curve when $\alpha=0$.
 
\section{Analysis \label{sec:analysis}}

\begin{table}
\begin{center}
\begin{tabular}{cccc}
  \hline
  \hline
  \hspace{5pt} $B_{ij}$ \hspace{5pt} & \hspace{5pt} $\ln{B_{ij}}$ \hspace{5pt}&\hspace{5pt} Evidence \hspace{5pt}\\
  \hline
  \hspace{5pt}$<2.5$  \hspace{5pt} &\hspace{5pt} $<1.1$ \hspace{5pt}&\hspace{5pt} Comparable       \\
  \hspace{5pt}$<20$ \hspace{5pt} &\hspace{5pt} $<3$   \hspace{5pt}&\hspace{5pt} Weak   \\
  \hspace{5pt}$<150$\hspace{5pt} &\hspace{5pt} $<5$   \hspace{5pt}&\hspace{5pt} Moderate     \\
  \hspace{5pt}$>150$\hspace{5pt} &\hspace{5pt} $>5$   \hspace{5pt}&\hspace{5pt} Strong\\
  \hline
\end{tabular}
\caption{The values of both the linear and the logarithmic Jeffreys' scale. \label{tab:Jef}} 
\end{center}
\end{table}

\begin{table}
\begin{centering}
\begin{tabular}{|c|c|c|c|}
\hline Models & $B$ & $\ln (B)$ & $ \chi^2$ \\
\hline Hellings-Downs & - & - & 10.25 \\
Genetic Algorithms & - & - & 2.83 \\
Ultralight vector DM & 0.29 & -1.24 & 6.39 \\
Spin-2 ultralight DM & 0.80 & -0.22 & 8.43 \\
Massive gravity & 1.51 & 0.41 & 6.91 \\
Non-Gaussianity of SGWB & 0.06 & -2.85 & 3.17 \\
\hline
\end{tabular}
\par
\end{centering}
\caption{The values of log Bayes evidences and the differences in $\chi^2$ for the different models, obtained using the NANOGrav data.\label{tab:models}}
\end{table}

\begin{table}
\begin{centering}
\begin{tabular}{|c|c|}
\hline Models & Best fit parameters  \\
\hline
Ultralight vector DM & $\mu=1\times 10^{-24}$ \\
\hline
Spin-2 ultralight DM & $m=4.4\times 10^{-24}$, $\alpha=5.5\times 10^{-6}$ \\
\hline
Massive gravity & $A$=0.73, $\Omega$=0.46 \\
\hline
Non-Gaussianity of SGWB & $\alpha=-0.99$ \\
\hline
\end{tabular}
\par
\end{centering}
\caption{The best-fit parameters for each model, obtained using the NANOGrav data.}\label{tab:bestfit}
\end{table}

\begin{figure*}[!thb]
\centering
\includegraphics[width = 0.9\textwidth]{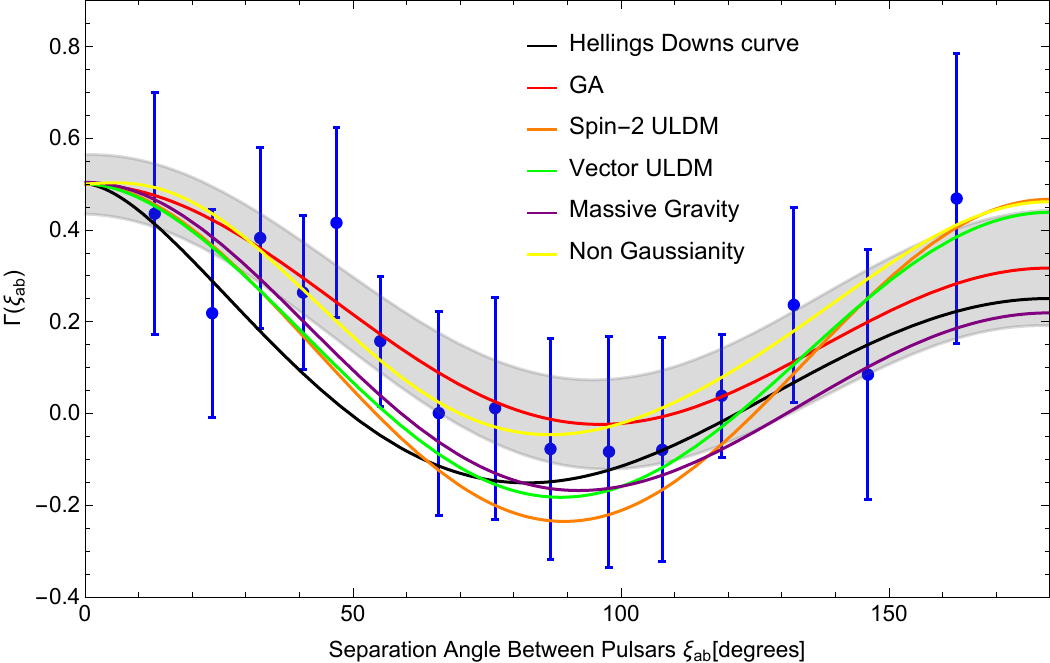}
\caption{Comparison of overlap reduction functions for different models. The red curve represents our GA best-fit to the data, with the gray band indicating its 1$\sigma$ error. The colored curves represent various models, alternative to the GR prediction, with best-fit parameters: the orange curve corresponds to $\Gamma_\mathrm{eff}(\xi)$ for the spin-2 ultralight DM model, the green curve represents the vector DM model, the purple curve corresponds to the massive gravity model, and the yellow curve illustrates the folded non-Gaussian component model. Finally, the black curve, shown for reference, represents the HD prediction, while the blue dots with error bars indicate the NANOGrav 15-year dataset. \label{fig:HD_GA}}
\end{figure*}

For our analysis, we use the reconstructed overlap reduction function obtained through the frequentist optimal statistic from the 15-year pulsar timing dataset collected by the NANOGrav collaboration~\cite{agazie2023nanograv}. The angular-separation–binned inter-pulsar correlations were constructed using 2,211 pairings within the 67-pulsar array. For that, the spectral index was fixed at $\gamma = 13/3$, and maximum-a-posteriori values obtained from a Bayesian inference analysis were assumed for the pulsar noise parameters and common-process amplitude. In Fig.~\ref{fig:HD_GA}, the dashed black line represents the theoretical HD correlation pattern, while the NANOGrav data points and error bars are shown in blue.

In order to compare the models we are considering, we need to perform Bayesian nested model comparison. Bayes factors for two nested models can be calculated using the Savage–Dickey density ratio~\cite{dickey1971weighted,verdinelli1995computing}; see also Ref.~\cite{mukherjee2006model} and Appendix \ref{sec:SD} for a full derivation. Assuming a Gaussian approximation of the likelihood, the Savage–Dickey formula for an extended model $M$ with $n$ parameters $\theta_i$ and flat priors $\Delta \theta_i$, compared to a simpler model $M'$ with parameters $n'$, where $n=n'+p$, is given by:  
\be
B = \left(\Pi_{i=n'}^{p}\Delta \theta_i\right)\,(2\pi)^{-p/2}\,|F_{pp}|^{1/2} \,\exp \left[-\frac12\left(\chi^{'2}_\mathrm{min}-\chi^{2}_\mathrm{min}\right)\right],
\ee 
where $F_{pp}$ is the marginalized $p\times p$ Fisher matrix of the extra $p$ parameters, while $\chi^{'2}_\mathrm{min}$ and $\chi^{2}_\mathrm{min}$ are the best-fit chi-squares of the $M'$ and $M$ respectively (see Appendix \ref{sec:SD} for more details). In our specific case, $M$ consists of four models: ultralight vector DM, spin-2 ultralight DM, massive gravity, and non-Gaussian component to SGWB, while $M'$ is the HD curve. Finally, we can also write the log-Bayes factor as 
\ba 
\ln B \simeq -\frac12\left(\chi^{'2}_\mathrm{min}-\chi^{2}_\mathrm{min}\right)+\ln |F_{pp}|^{1/2} + \ln \left[\frac{\Pi_{i=n'}^{p}\Delta \theta_i}{\,(2\pi)^{p/2}}\right].\nn \\ \label{eq:SD2}
\ea

To assess the strength of evidence supporting or opposing a model in a comparison between two models using the Bayes factor $B$ (which represents the ratio of evidences), the updated Jeffreys' scale can be used~\cite{trotta2008bayes}. Based on this scale, if $|\ln B|<1.1$, the models are comparable, with neither one being distinctly preferred. When $1.1<|\ln B|<3$ there is weak evidence favoring one model. For $3<|\ln B|<5$ the evidence grows moderate, and if $|\ln B|>5$, there is strong support for one model over the other. For convenience, we provide the specific values of both the linear and logarithmic Jeffreys' scales in Table \ref{tab:Jef}.

In Table~\ref{tab:models}, we present the Bayes and log Bayes evidences along with the differences $\chi^2$ for different models using the NANOGrav 15-year data set, contrasted to the regular HD curve. For the HD curve we obtain a fit to the data of  $\chi^2=10.25$. We also fitted the PTA data through a particular machine learning approach, named Genetic Algorithms (GA). Our GA algorithm is a non-parametric approach to describing the data, searching for the best-fit functional form to represent it. See Appendix~\ref{sec:GA} for a brief description of this algorithm. Since the GA do not have free parameters, we can only compute the $\chi^2$. This model independent analysis has been applied to the data for comparison reasons, since non-parametric methods offer greater flexibility and can identify a broader array of patterns in the data, minimizing theoretical assumptions.

As expected, Table~\ref{tab:models} shows that the GA provides the best fit to the data in terms of the $\chi^2$ value. For the ultralight vector DM model our analysis gives a log Bayes ratio of $\ln (B) = -1.24$, suggesting a weak evidence in favor of this model and against the regular HD curve. Second, the spin-2 ultralight DM model yields a log Bayes ratio of $\ln (B) = -0.22$, indicating that the HD curve and this model are comparable according to Jeffrey's scale. Third, in the case of the massive gravity model we find that $\ln (B)=0.41$, suggesting that this model and the HD curve are statistically equivalent, with neither being distinctly preferred. Finally, our fourth model in consideration is the implementation of a non-gaussianity parameter of SGWB where $\ln (B)=-2.85$ resulting in weak evidence in favor of this model and against the HD curve. We found that $\alpha=-0.99$ produced the best fit to the data. However, note that the authors of Ref.~\cite{jiang2024search} present a different claim, stating that a negative $\alpha$ is strongly disfavored.

This discrepancy is likely due to differences in the methods used for normalization. 
We also find $\ln (B)=-2.19$ for $\alpha=10$, showing that this parameter region is statistically in weak evidence in favor of this model.

In Fig.~\ref{fig:HD_GA}, we present the different functional forms of the overlap reduction function, $\Gamma_\mathrm{eff}(\xi)$, for each model, using the best-fit parameter values listed in Table~\ref{tab:bestfit}. 
The red curve represents our GA best-fit to the data, with the gray band indicating its 1$\sigma$ error. 
For reference, the black curve represents the HD curve, while the blue dots with error bars correspond to the NANOGrav 15-year dataset.
Interestingly, the model-independent fit by GA exhibits a slight deviation from the standard HD curve, exceeding 1$\sigma$ in certain angular ranges, although higher-quality data will be needed to provide stronger evidence for a claim. We must also take into account the intrinsic variance of the HD curve~\cite{allen2023variance,allen2023hellings}. The colored curves depict the best fit for each model: the orange curve corresponds to $\Gamma_\mathrm{eff}(\xi)$ for the spin-2 ultralight DM model, the green curve represents the vector DM model, the purple curve corresponds to the massive gravity model, and the yellow curve illustrates the folded non-Gaussian component model, which is the preferred model among the four considered. We can see that the folded non-Gaussian component model has the most similar shape function than the GA reconstruction, lying in between the gray error band.  

\section{Conclusions  \label{sec:conclusions}}

In this study, we have performed a comparative Bayesian analysis using the NANOGrav 15-year dataset to explore four distinct models that modify the overlap reduction function of the SGWB. These models include ultralight vector DM, spin-2 ultralight DM, massive gravity, and a folded non-Gaussian component to SGWB. The objective was to assess their ability to reproduce the observed correlations between pulsars, particularly the well-known HD curve, which serves as a key signature of an isotropic SGWB in the framework of GR. 

Our analysis yields several key findings:
For the spin-2 Ultralight DM and massive gravity model
both of them exhibit Bayesian evidence suggesting statistical equivalence with the HD curve. Specifically, the Bayesian log ratio for the spin-2 Ultralight DM model is $\ln (B)=-0.22$, and for the massive gravity model it is $\ln (B)=0.41$. These results indicate that neither of these models is strongly preferred over the traditional HD curve, implying that modifications from these DM models do not significantly improve the fit to the data. In contrast, the data shows weak evidence in favor of the ultralight vector DM model and the folded non-Gaussian component, with a log Bayes factor of $\ln (B)=-1.24$ and $\ln (B)=-2.85$ respectively. This suggests that including an ultralight bosonic field or a folded non-Gaussian component provide a better description of the observed data, though the evidence is not conclusive. 

Overall, our results demonstrate that while modifications to the standard HD curve offer alternative explanations for the observed data, no single model provides definitive evidence of a departure from the HD curve. The weak statistical preferences observed across multiple models suggest that current data is not yet sufficient to decisively favor any of these theoretical extensions. Future datasets with improved signal-to-noise ratios may help to clarify the validity of these models and provide valuable insights into fundamental physics beyond the current framework of cosmology.

\section*{Acknowledgements}
The authors also acknowledge support from the research project PID2021-123012NB-C43 and the Spanish Research Agency (Agencia Estatal de Investigaci\'on) through the Grant IFT Centro de Excelencia Severo Ochoa No CEX2020-001007-S, funded by MCIN/AEI/10.13039/501100011033. SK's research is supported by the Spanish Attraccion de Talento contract no. 2019-T1/TIC-13177 granted by the Comunidad de Madrid. SK is also partly supported by the I+D grant PID2020-118159GA-C42 funded by MCIN/AEI/10.13039/501100011033, the Consolidaci\'on Investigadora 2022 grant CNS2022-135211, and Japan Society for the Promotion of Science (JSPS) KAKENHI Grant no. 20H01899, 20H05853, and 23H00110.

{\bf Numerical Analysis Files}: The Genetic Algorithm code used by the authors in this paper can be found at \href{https://github.com/RubenArjona}{https://github.com/RubenArjona}.\\

\begin{appendix} 

\section{Genetic Algorithms \label{sec:GA}}
Here, we provide a brief overview of the genetic algorithms (GAs) and their integration into our analytical framework. GAs have been extensively used in cosmology and have demonstrated effectiveness in reconstructing various datasets across diverse contexts. For more detailed discussions, readers can refer to several relevant studies~\cite{Bogdanos:2009ib,Nesseris:2010ep, Nesseris:2012tt,Nesseris:2013bia,Sapone:2014nna,Arjona:2020doi,Arjona:2020kco,Arjona:2019fwb,Arjona:2021hmg,Arjona:2020skf,Arjona:2020axn,aizpuru2021machine,arjona2024probing,arjona2022testing,arjona2021complementary,arjona2024swampland}. Outside of cosmology, genetic algorithms (GAs) are also applied in particle physics~\cite{Abel:2018ekz,Allanach:2004my,Akrami:2009hp}, astrophysics~\cite{wahde2001determination,Rajpaul:2012wu,Ho:2019zap} 
as well as pulsar timing parameter estimation~\cite{Petiteau:2012zq}.  Additional uses of symbolic regression techniques in physics and cosmology can be found in Refs.~\cite{Udrescu:2019mnk,Setyawati:2019xzw,vaddireddy2019feature,Liao:2019qoc,Belgacem:2019zzu,Li:2019kdj,Bernardini:2019bmd,Gomez-Valent:2019lny}.

In our approach, we model the desired NANOGrav 15-year data set as chromosomes, similar to the composite structure in genetics, with individual features acting as 'genes.' A fitness function, comparable to a loss function in machine learning, links the chromosomes to the properties of the target solution, allowing us to assess how closely a candidate solution aligns with the available data. In this case, the fitness function is defined as the sum of $\chi^2$ values from various likelihoods. By employing GAs, we aim to optimize feature selection and iteratively refine the solution space to find the configurations that best match the data. The basic steps of a GA can be summarized as follows:
\begin{itemize}
    \item Initialization: A population of potential solutions is randomly generated, with each individual representing a possible solution to the optimization problem.
    \item Fitness Evaluation: Each individual’s fitness is measured using a fitness function that evaluates how well they solve the problem.
    \item Selection: The best-performing individuals are chosen to create the next generation, with the probability of selection based on their fitness scores.
    \item  Crossover: The selected individuals combine genetic information to create a new generation, with offspring formed by mixing traits from each parent.
\item Mutation: A small proportion of the offspring undergo random mutations to introduce genetic diversity into the population.
\item Termination: The algorithm concludes when a set number of generations is reached or when a solution meets predefined fitness criteria.
\end{itemize}
After the GA code terminates, it produces an analytic best-fit function, which describes the data and can be used for further analyses.

\section{Overlap reduction function expressions \label{sec:func}}
We present a detailed derivation of the overlap reduction functions for the four models considered in this paper.

\subsection{Ultralight vector dark matter}
\label{app:vector}
The overlap reduction function $\Gamma_{\mathrm{eff}}$ for vector DM, within the frequency band encompassing $2\pi f = 2\mu$, where $\mu$ denotes the DM mass, can be expressed as
\begin{equation}
\small{
\Gamma_{\mathrm{eff}}(\xi,\mu)=\frac{\Phi_{\mathrm{GW}}(\mu / \pi)}{\Phi_{\mathrm{GW}}(\mu / \pi)+\Phi_{\mathrm{DM}}}\left[\Gamma_{\mathrm{HD}}(\xi)+\frac{\Phi_{\mathrm{DM}}}{\Phi_{\mathrm{GW}}(\mu / \pi)} \Gamma_{\mathrm{DM}}(\xi)\right]} \,.
\end{equation}
The normalization is chosen such that $\Gamma_{\mathrm{eff}}(0)=1/2$, with $\Gamma_{\mathrm{HD}}(\xi)$ representing the standard HD pattern produced by a SGWB. The function $\Phi_{\mathrm{GW}}(\mu / \pi)$ at $f=\mu/\pi$ is estimated to be~\cite{omiya2023hellings}:
\begin{equation}
\Phi_{\mathrm{GW}}(\mu / \pi) \sim 5 \times 10^{-34} \mathrm{yr}^2\left(\frac{\mu}{10^{-22} \mathrm{eV}}\right)^{-13 / 3}\left(\frac{15 \mathrm{yr}}{T_{\mathrm{obs}}}\right),
\end{equation}
where $T_{\rm obs}$ is the observation time. The function $\Phi_{\mathrm{DM}}$ is assumed under the condition that DM consists entirely of ultralight DM and disregarding any stochastic effects, and is defined as
\begin{equation}
\Phi_{\mathrm{DM}} \sim 7 \times 10^{-37} \mathrm{yr}^2\left(\frac{\rho_{\mathrm{DM}}}{0.4 \mathrm{GeV} \cdot \mathrm{~cm}^{-3}}\right)^2\left(\frac{10^{-22} \mathrm{eV}}{\mu}\right)^6,
\end{equation}
where $\rho_\mathrm{DM}$ is the DM density and taken to be $\rho_{\mathrm{DM}}=0.4 \mathrm{GeV} \cdot \mathrm{cm}^{-3}$.

\subsection{Spin-2 ultralight dark matter}
\label{app:spin2}
The overlap reduction function $\Gamma_{\mathrm{eff}}$ of spin-2 ultralight DM at frequency $f_m=m/2\pi$, where $m$ is the mass of the DM particle, can be written as~\cite{cai2024angular}
\begin{equation}
\begin{aligned}
\Gamma_{\mathrm{eff}}(\xi,m,\alpha)= & \frac{\Phi_{\mathrm{GW}}(m / 2 \pi)}{\Phi_{\mathrm{GW}}(m / 2 \pi)+\Phi_{\mathrm{DM}}(\alpha)} \Gamma_{\mathrm{HD}}(\xi)+ \\
& \frac{\Phi_{\mathrm{DM}}(\alpha)}{\Phi_{\mathrm{GW}}(m / 2 \pi)+\Phi_{\mathrm{DM}}(\alpha)} \Gamma_{\mathrm{DM}}(\xi)
\end{aligned}
\end{equation}
where $\alpha$ is the parameter that corresponds to the coupling between the ultralight DM field and matter and the normalization is selected such that $\Gamma_{\mathrm{eff}}(0)=1/2$ and $\Gamma_{\mathrm{HD}}(\xi)$ is the HD pattern due to a SGWB. 

The amplitude of the correlation due to the SGWB $\Phi_{\mathrm{GW}}(m / 2 \pi)$ is given by
\begin{equation}
\Phi_{\mathrm{GW}}(m / 2 \pi) \sim 1 \times 10^{-32} \mathrm{yr}^2\left(\frac{m}{10^{-22} \mathrm{eV}}\right)^{-\frac{13}{3}}\left(\frac{15 \mathrm{yr}}{T_{\mathrm{obs}}}\right) \,.
\end{equation}
The amplitude of the correlation due to the ultralight DM can be written as
\begin{equation}
\begin{aligned}
\Phi_{\mathrm{DM}} \sim  6 & \times 10^{-33} \mathrm{yr}^2\left(\frac{\rho_{\mathrm{DM}}}{0.4 \mathrm{GeV} / \mathrm{cm}^3}\right) \\
& \times\left(\frac{\alpha}{10^{-6}}\right)^2\left(\frac{m}{10^{-22} \mathrm{eV}}\right)^{-4},
\end{aligned}
\end{equation}
where $\rho_{\mathrm{DM}}=0.4 \mathrm{GeV} \cdot \mathrm{cm}^{-3}$.\\

\subsection{Massive gravity}
\label{app:massive}
For massive gravity model, the effective overlap reduction function can be given by summing the contributions from tensor, vector, and scalar modes,
\begin{equation}
\Gamma_{\mathrm{eff}}(\xi,A, \Omega)=\Gamma_T(\xi,A)+\Omega\cdot \Gamma_V(\xi,A)+\Omega \cdot \Gamma_S(\xi,A).
\end{equation}

The overlap reduction function of the tensor modes is given by
\begin{widetext}
\begin{equation}
\begin{aligned}
\Gamma_{0, T}=\frac{-\pi}{6 A^5} & \frac{\beta_T}{4}\left[4 A\left(-3+\left(-6+5 A^2\right) \cos \xi\right)+12\left(1+\cos \xi+A^2(1-3 \cos \xi)\right) \ln \frac{1+A}{1-A}\right. \\
& \left.+\frac{3\left(1+2 A^2(1-2 \cos \xi)-A^4\left(1-2 \cos^2 \xi\right)\right) \ln L_1}{\sqrt{(1-\cos \xi)\left(2-A^2(1+\cos \xi)\right)}}\right],
\end{aligned}
\end{equation}
where $A=\frac{|\boldsymbol{k}|}{k_0}$, $\beta_T=\frac{3}{4\pi}$ which is a normalization factor introduced to impose $\Gamma(|f|)=1$ for coincident, co-aligned detectors~\cite{liang2021detecting}, and
\begin{equation}\label{eq:L1}
L_1 \equiv \frac{\left[1+2 A^2(1-2 \cos \xi)-A^4\left(1-2 \cos^2 \xi\right)-2 A\left(1-A^2 \cos \xi\right) \sqrt{(1-\cos \xi)\left(2-A^2(1+\cos \xi)\right)}\right]^2}{\left(1-A^2\right)^4}.
\end{equation}
For the vector modes, the overlap reduction function is defined as
\begin{equation}
\begin{aligned}
& \Gamma_{0, V}=\frac{\beta_V}{8} \frac{k_0^2}{m^2} \frac{8 \pi}{A^5}\left[\frac{A}{3}\left(A^6 \cos \xi-2 A^4(5 \cos \xi+3)+2 A^2(11 \cos \xi+6)-6(2 \cos \xi+1)\right)\right. \\
+ & \left.\left(A^2-1\right)^2\left(\left(A^2-2\right) \cos \xi-2\right) \ln \left(\frac{1-A}{1+A}\right)+\frac{\left(A^2-1\right)^2\left(1-A^2 \cos \xi\right) \ln L_1}{2 \sqrt{(1-\cos \xi)\left(2-A^2(1+\cos \xi)\right)}}\right],
\end{aligned}
\end{equation}
where $\beta_V=\frac{3}{4\pi}$, $\frac{k_0^2}{m^2} =\frac{1}{1-A^2}$ and $L_1$ is defined in Eq.~\eqref{eq:L1}. Finally, for the scalar mode, the overlap reduction function is defined as
\begin{equation}
\begin{aligned}
\Gamma_{0, S}=\frac{\beta_S}{4} \frac{\pi}{6 A^5} & \left[4 A\left(9+4 A^4+18 \cos \xi-3 A^2(4+5 \cos \xi)\right)-12\left(1-A^2\right)\left(2 A^2-3-3 \cos \xi\right) \ln \frac{1-A}{1+A}\right. \\
& \left.-\frac{9\left(1-A^2\right)^2 \ln L_1}{\sqrt{(1-\cos \xi)\left(2-A^2(1+\cos \xi)\right)}}\right],
\end{aligned}
\end{equation}
\end{widetext}
where $\beta_S=\frac{3}{4\pi}$ and $L_1$ is defined in Eq.~\eqref{eq:L1}. 

\subsection{non-Gaussian component to SGWB}
\label{app:nonG}
For the non-Gaussian component model, the overlap reduction function is given by
\small{
\begin{equation}
\begin{aligned}
&\Gamma\left(\xi_{a b}\right) \propto  \\
& \sum_{A=+, \times}\int_{S^2} \mathrm{~d} \hat{\mathbf{n}} \frac{\bar{E}_a^A \bar{E}_b^A}{\left(1+\hat{\mathbf{n}}_a \cdot \hat{\mathbf{n}}\right)\left(1+\hat{\mathbf{n}}_b \cdot \hat{\mathbf{n}}\right)} +\\
& \frac{+4 \alpha\left(\frac{9}{16} \bar{E}_a^A \bar{E}_a^A \bar{E}_b^A \bar{E}_b^A+\frac{5}{8}\left(\bar{E}_a^A \bar{E}_a^A \bar{E}_a^A \bar{E}_b^A+\bar{E}_a^A \bar{E}_b^A \bar{E}_b^A \bar{E}_b^A\right)\right)}{\left(1+\hat{\mathbf{n}}_a \cdot \hat{\mathbf{n}}\right)\left(1+\hat{\mathbf{n}}_b \cdot \hat{\mathbf{n}}\right)},
\end{aligned}
\end{equation}
}
\normalsize{}
\noindent where $\bar{E}_a^A=\mathbf{e}_{i j}^A \hat{\mathbf{n}}_a^i \hat{\mathbf{n}}_a^j$, and $\mathbf{e}_{i j}^A$ is the basis for the two kinds of symmetric polarization tensors $\left(+, \times\right)$.
We represent the GW direction along the spatial coordinates in a Cartesian system using $\hat{n}$, which corresponds to $\left(\hat{x},\hat{y},\hat{z}\right)$. Additionally, we define two unit vectors, $\hat{u}$ and $\hat{v}$, that are orthogonal to $\hat{n}$, see also~\cite{tasinato2022gravitational}:
\begin{equation}
    \begin{array}{rcl}\hat u&=&\frac{\hat n \times \hat z}{|\hat n \times \hat z|} ,\\\hat v&=&\frac{\hat n \times \hat u}{|\hat n \times \hat u|} .\end{array}
\end{equation}
These quantities can also be represented in spherical coordinates as
\begin{equation}
    \begin{array}{rcl}\hat{n}&=&(\sin\theta\cos\phi,\sin\theta\sin\phi, \cos\theta) ,\\\hat{u}&=&(\cos\theta\cos\phi, \cos\theta\sin\phi, -\sin\theta) ,\\\hat{v}&=&(\sin\phi, -\cos\phi, 0) .\end{array}
\end{equation}
The vectors  $\hat{u}$ and  $\hat{v}$ are not the only unit vectors orthogonal to  $\hat{n}$. In a more general case, $\hat{u}$ and  $\hat{v}$  can be rotated around $\hat{n}$ by an angle $\psi$:
\begin{equation}
    \begin{aligned}\hat{u}'&&=&\cos\psi \hat{u}+\sin\psi \hat{v} ,\\\hat{v}'&&=&-\sin\psi \hat{u}+\cos\psi \hat{v} .\end{aligned}
\end{equation}
Then we can write the two kinds of polarization tensors as
\begin{equation}
\begin{array}{rcl}
    \mathbf{e}_{ij}^{(+)} &=& \hat{u}'_i\hat{u}'_j-\hat{v}'_i\hat{v}'_j ,\\
    \mathbf{e}_{ij}^{(\times)} &=& \hat{u}'_i\hat{v}'_j-\hat{v}'_i\hat{u}'_j ,
    \end{array}
\end{equation}
where later we have to average over this angle $\psi$ since observables should not depend on it. Finally, to have fully defined the vector $\bar{E}_a^A$ and $\bar{E}_b^A$ we have that
\begin{equation}
    \begin{array}{rcl}\hat{n}_a&=&(0,0,1) ,\\\hat{n}_b&=&(\sin\xi, 0, \cos\xi).\end{array}
\end{equation}
To reproduce Figure 1 of Ref.~\cite{jiang2024search} we have to solve the following integrals
\small{
\begin{equation}
\begin{aligned}
&\Gamma\left(\xi\right) \propto  \\
& \sum_{A=+, \times}\frac{3}{32\pi}\int^{\pi}_{0}\int^{2\pi}_{0}\int^{\pi}_{0} \sin\theta d\psi d\phi d\theta \left(\frac{\bar{E}_a^A \bar{E}_b^A}{\left(1+\hat{\mathbf{n}}_a \cdot \hat{\mathbf{n}}\right)\left(1+\hat{\mathbf{n}}_b \cdot \hat{\mathbf{n}}\right)} +\right.\\
& \left.\frac{+4 \alpha\left(\frac{9}{16} \bar{E}_a^A \bar{E}_a^A \bar{E}_b^A \bar{E}_b^A+\frac{5}{8}\left(\bar{E}_a^A \bar{E}_a^A \bar{E}_a^A \bar{E}_b^A+\bar{E}_a^A \bar{E}_b^A \bar{E}_b^A \bar{E}_b^A\right)\right)}{\left(1+\hat{\mathbf{n}}_a \cdot \hat{\mathbf{n}}\right)\left(1+\hat{\mathbf{n}}_b \cdot \hat{\mathbf{n}}\right)}\right),
\end{aligned}
\end{equation}
}
\normalsize{}
and finally normalize it so that $\Gamma\left(\xi=0\right)=1/2$.

\section{Bayes factors and Savage-Dickey formula \label{sec:SD}}

Here we provide a brief derivation of the Savage-Dickey formula, as applied in our case, i.e. comparing a simple model with a more complicated model, in the case where they are nested. To keep the notation simple but general, we consider a simpler model $M'$ with $n'$ parameters denoted by $\theta_i$, $i\in[1,n']$ and a more general model $M$ with $n=n'+p$ parameters $\theta_i$, where $i\in[1,n]$, so that $n'<n$. 

Then, assuming Gaussianity, the $\chi^2$ for both cases can be written as 
\ba 
\chi^{2}(\theta)=\chi^{2}_\mathrm{min}+\left(\theta-\theta_\mathrm{min} \right)_i\,F_{ij} \,\left(\theta-\theta_\mathrm{min} \right)_j+\ldots,~~ 
\ea 
where $F_{ij}$ is the Fisher matrix of the parameters and the unnornalized likelihood is of the form
\be 
\mathcal{L} = \exp \left[-\chi^{2}(\theta)/2\right].
\ee 
The evidence is then the integral over all the parameters. Assuming flat priors in some range $\Delta \theta_i$, which we can assume they are much larger than the width of the likelihood so that the integration limits can be extended to infinity, we have for the $M'$ model:
\ba 
E_{M'} &=& \int \left(\Pi_i^{n'} \Delta \theta_i^{-1}\right)\,\exp \left[-\chi^{2}(\theta)/2\right]\, \mathrm{d}^{n'}\theta \nn \\
&\simeq& \left(\Pi_i^{n'} \Delta \theta_i^{-1}\right)\,(2\pi)^{n'/2}\,\exp \left[-\chi^{'2}_\mathrm{min}/2\right]\, |F'|^{-1/2},\nn\\
\ea 
and similarly for the $M$ model:
\ba 
E_{M}\simeq \left(\Pi_i^{n} \Delta \theta_i^{-1}\right)\,(2\pi)^{n/2}\,\exp \left[-\chi^{2}_\mathrm{min}/2\right]\, |F|^{-1/2},~~~~~
\ea 
where in both cases $F$ and $F'$ are the Fisher matrices of the parameters for the two models. 

The Bayes ratio between the two models $M'$ and $M$ is then 
\ba 
B &\equiv& \frac{E_{M'}}{E_{M}} \nn \\
&\simeq& \frac{\left(\Pi_i^{n'} \Delta \theta_i^{-1}\right)\,(2\pi)^{n'/2}\,\exp \left[-\chi^{'2}_\mathrm{min}/2\right]\, |F'|^{-1/2}}{\left(\Pi_i^{n} \Delta \theta_i^{-1}\right)\,(2\pi)^{n/2}\,\exp \left[-\chi^{2}_\mathrm{min}/2\right]\, |F|^{-1/2}} \nn \\
&=& \left(\Pi_{i=n'}^{p}\Delta \theta_i\right)\,(2\pi)^{-p/2}\,|F_{pp}|^{1/2} \,\exp \left[-\frac12\left(\chi^{'2}_\mathrm{min}-\chi^{2}_\mathrm{min}\right)\right],\nn \\
\label{eq:SD1}\ea 
where we used the fact that $n=n'+p$ and that in the case of the nested models, the Fisher matrix of the bigger model $F$ with dimensions $n\times n$ is a block matrix with respect to the smaller model $F'$ with dimensions $n'\times n'$ and $n>n'$. Thus, schematically we have 
\be F_{nn}=\left(
\begin{array}{cc}
   F^{'}_{n'n'}  & F_{n'p} \\
   F_{pn'}  &  F_{pp}
\end{array}\right),
\ee 
where the subscripts indicate the range of the indices. Then, we have the identity,
\ba 
|F| &=& |F_{pp}|\, |F'-F_{n'p} \, F_{pp}^{-1} \, F_{pn'}| \nn \\
&\simeq& |F_{pp}|\, |F'|,
\ea 
where in the second line we neglected the correlations between the new parameters and the old ones. Thus, we can write the log-Bayes factor as 
\ba 
\ln B \simeq -\frac12\left(\chi^{'2}_\mathrm{min}-\chi^{2}_\mathrm{min}\right)+\ln |F_{pp}|^{1/2} + \ln \left[\frac{\Pi_{i=n'}^{p}\Delta \theta_i}{\,(2\pi)^{p/2}}\right].\nn \\ \label{eq:SD2}
\ea 

Overall, if $\ln B>0$ there is some evidence in favor of model $M'$ (the simpler model), otherwise if $\ln B<0$ there is some evidence in favor of model $M$ (the bigger model), which can interpreted with the Jeffreys scale.

Finally, it should be noted that Eqs.~\eqref{eq:SD1} and \eqref{eq:SD2} should be used and interpreted with care, as they make several assumptions:
\begin{itemize}
    \item The likelihood is close to a Gaussian.
    \item The new parameters $p$ are very weakly correlated with the old ones, i.e. $F_{n'p} \simeq 0$.
    \item The flat priors are much larger than the width of the likelihood, so that the integration can be extended to infinity.
\end{itemize}

\end{appendix}

\bibliography{WL}

\end{document}